# R²U3D: Recurrent Residual 3D U-Net for Lung Segmentation

**Dhaval D. Kadia[1],[*], Member, IEEE, Md Zahangir Alom[3], Ranga Burada[4], Tam V. Nguyen[1], Senior Member, IEEE, and Vijayan K. Asari[2], Senior Member, IEEE**

[1] Department of Computer Science, University of Dayton, Dayton, OH 45469 USA
[2] Department of Electrical and Computer Engineering, University of Dayton, Dayton, OH 45469 USA
[3] St. Jude Children's Research Hospital, Memphis, TN 38105 USA
[4] Microsoft, Redmond, WA 98052 USA

[*]Corresponding author: Dhaval D. Kadia (e-mail: kadiad1@udayton.edu).

This paragraph of the first footnote will contain support information, including sponsor and financial support acknowledgment.

**ABSTRACT** 3D lung segmentation is essential since it processes the volumetric information of the lungs, removes the unnecessary areas of the scan, and segments the actual area of the lungs in a 3D volume. Recently, the deep learning model, such as U-Net outperforms other network architectures for biomedical image segmentation. In this paper, we propose a novel model, namely, Recurrent Residual 3D U-Net (**R²U3D**), for the 3D lung segmentation task. In particular, the proposed model integrates 3D convolution into the Recurrent Residual Neural Network based on U-Net. It helps learn spatial dependencies in 3D and increases the propagation of 3D volumetric information. The proposed R²U3D network is trained on the publicly available dataset LUNA16 and it achieves state-of-the-art performance on both LUNA16 (testing set) and VESSEL12 dataset. In addition, we show that training the R²U3D model with a smaller number of CT scans, i.e., 100 scans, without applying data augmentation achieves an outstanding result in terms of Soft Dice Similarity Coefficient (Soft-DSC) of 0.9920.

## I. INTRODUCTION

Lung cancer is considered the second most common cancer type in both men and women [1]. The lung cancer patient is more likely to be successfully treated if it is found at an earlier stage, and before it has spread. Many patients having lung cancer report many kinds of delays in the diagnosis. The patients waited a median of 21 days before visiting a doctor and more than 22 days to complete the investigations. The median wait to start the treatment once the patients were seen at the cancer center was ten days. The total time from the development of the first symptoms to starting treatment was 138 days [2]. This affects the survival possibilities of the patients. Lung cancer screening using the low-dose CT is used to treat the patient to reduce lung cancer mortality.

The lung segmentation is important because it gives the volumetric information of the lungs. It is challenging because the lungs have irregular shapes, sizes, low contrasts, and complex boundaries [16]. Moreover, lung segmentation removes the unnecessary areas of the CT scan and segments the lungs' actual area, where the attention is much essential. Lung segmentation prevents computer program to process irrelevant volumetric data that can produce false positives and leads to the erroneous diagnosis. Additionally, it can be considered as a necessary preprocessing for different lung disease analysis such as lung nodule detection or segmentation, pulmonary embolism (PE) diagnosis, Acute Respiratory Distress Syndrome (ARDS), and pneumothorax analysis [17][19][20].

Lung segmentation helps to save annotation time. Particularly for 3D segmentation applications, annotation is time-consuming, and such segmentation application can produce the segmentation that can be corrected with some additional efforts.

Traditional methods such as thresholding, edge tracking, region growing, contrast, and neighborhood homogeneity are applied for lung segmentation, but these methods do not give promising results when the CT scan of lungs is infected or has high attenuation patterns. They use edge detection filters and other mathematical operations and algorithms. These methods have advantages if the data are less diverse, and domain knowledge is applied correctly, and they give accurate results. Using the patch-based approach limits feature extraction by the number of patches, and that affects learning. The texture-based methods addressed such situations but gave poor results when some abnormalities were in the peripheral lung. Using traditional methods, the lung segmentation of a 3D CT scan can also be two-dimensional by applying 2D segmentation on each slice. The study shows that inter-slice smoothness is significantly smoother in 3D segmentation than 2D segmentation [15][29].

In 1959, Arthur Samuel described machine learning as the "field of study that gives computers the ability to learn without being explicitly programmed" [30]. Deep learning (DL) is a subfield of machine learning, a field within artificial intelligence (AI). DL consists of a multilayer neural network that extracts features in more depth. The convolutional neural network (CNN) is the most powerful architecture in deep learning. CNN correlates nearby pixels of an image and produces different outputs using respective sets of weights. These sets of weights extract the features from an image. Repeating this process further gives us the features of an image. The initial stages of feature extraction give low-level features, and since the further stages extract the features from the previous or earlier stages, the later stages give high-level features. Low-level features help to correlate and understand small details, and high-level features represent the big picture or summarization of previous low-level features. A fully connected network is very bulky, whereas CNN has less trainable spatial feature extraction parameters. In recent years, the deep convolutional neural networks are vital and outperforming state of the art in feature extraction, visual recognition, and object segmentation. The deep neural networks contain millions of parameters to solve complex problems, and hence, it is quite necessary to have the right data in the proper format and enough amount to train the parameters. In less data availability, it is necessary to discover the neural network architecture that can be trained using less amount of training data.

Traditional machine learning applications use techniques like support vector machines (SVMs) and random forests (RF). An issue with these approaches is that it requires collective efforts of field experts to approach useful features. Its optimization is time costly, and features are domain or problem-specific. Applicability of the same features among different domains is not always possible. Comparing to the traditional machine learning methods, deep learning has numerous advantages. Deep learning techniques learn useful features and do not require handcrafted features. Using transfer learning, the features learned from one dataset can be used to learn new features from different datasets. This gives importance to pre-trained deep learning systems trained on large datasets and sophisticated computational resources. They can be made available to the public to apply it to their applications. 3D convolutions are playing an essential role in spatial feature extraction in three-dimensions. Having fewer training data can be solved by training a 3D convolutional neural network on 3D patches of available data. This increases the training samples, and data augmentation helps further model generalization. Computer-aided diagnosis (CADx) requires sophisticated tools. It requires a deep neural network to learn complex features. Less training data will not let the deep neural network learn diverse features. So, instead of training a deep neural network from scratch, transfer learning can be applied, where the pre-trained model is trained on another dataset with enough diverse data. A significant transfer learning application is to use it for fine-tuning by freezing the initial layers of pre-trained convolutional neural networks and training the later layers. It works because high-level features differ among different datasets more than low-level features. It means better learned low-level feature extraction can be used to learn high-level features on a different dataset. During this process, the DL architecture remains the same; only weight gets updated. Another application of transfer learning is as a neural network weight initialization step. It helps the neural network to converge faster than other kinds of initialization approaches. Data augmentation generates new samples that increase the diversity in data points. Using such generated data for training reduces the probability of overfitting, and it overcomes the issue of the unbalanced dataset and helps generalize the neural network for testing dataset [14][21].

Deep learning is also applicable for photoacoustic tomography artifact removal [24]. This paper uses a Fully Dense Unet (FD-Unet) for removing artifacts. DL can be used to design an annotation tool, and it is more helpful for multi-dimensional data like 3D CT scans or such time-series data. It can help medical professionals to estimate the initial annotations and make further corrections [25]. Additionally, the corrections can train the DL model to get better for the next use. D-UNet discusses the problems of computational resources for 3D CNN and demonstrates the combined neural network of both 2D and 3D CNN for chronic stroke lesion segmentation [31]. It uses four slices as both 2D and 3D context to apply them to its neural network and achieves better results while combining 3D CNN. AUNet proposed an attention-guided dense-upsampling network for an alternative to deconvolution commonly used for the upsampling [32]. It explained that the deconvolution was not as effective as bilinear upsampling for their application of breast masses segmentation in mammograms. Research [33] proposes a multidimensional region-based fully convolutional neural network and combines three views of 3D CT scan to give the possible shape of the detected nodule and its classification as malignant or benign. X-Net was developed to effectively extract features with fewer trainable parameters using depthwise separable convolution for brain stroke lesion segmentation [34]. It also designed Feature Similarity Module to extract a wide range of position-sensitive contextual information. Thus, having a vast data dimension and given computational resources, it is challenging to develop 3D CNN for achieving excellent performance. Furthermore, we propose our methods to overcome these problems and fulfill expectations.

The artificial intelligence algorithms can be called trained algorithms, and they are becoming more complex and sophisticated to solve complicated problems. This requires algorithmic regulation to systemically review them to prevent unexpected harm without constraining the innovation. It is essential to know how deep learning algorithms learn and reason from their learning. It is required to know the metrics of algorithmic responsibility and how it can be traced. Human responsibility also plays a significant role in designing, improving, and maintaining the algorithm [26].

Deep neural networks have limitations to represent the learned knowledge to perform the assigned task explicitly. In such an environment, medical diagnostic tools need to be explainable, predictable, understandable, and transparent. This helps medical professionals, regulators, and patient's confidence and trust to understand how AI systems can be an integral part of routine diagnosis. Demonstrating the domain-specific features that help predict the output initially helps give an overview of a broader picture. Explainability is a necessary tool towards a trustworthy and ethical solution that is safe to use and have fairness in various aspects. It can be demonstrated using different approaches. Local interpretable methods give the reasoning for a single prediction, and the global methods give the abstract knowledge about the model, according to data [27]. The detailed analysis to understand the importance of a neuron is not limited by knowing the activation function's characteristics but necessitates its background learning process [28].

The physicians experience an increasing number of complex multi-dimensional visual readings, and this necessitates speed up clinical workflow with the help of deep learning and the technologies on top of that. While considering AI in medical imaging, we anticipate collaboratively using such technologies with physicians to decrease their burden, rather than replacing them.

The rest of this paper is organized as follows — section II reviews the related work. Section III discusses details of the proposed framework. Section IV reports the experimental results, and section V presents the evaluation of current research. Finally, Section VI concludes and paves the way to future work.

## II. RELATED WORK

The current progress in the deep learning algorithms and available machine learning architectures provide neural networks to perform complex feature extraction [8]. Deep learning algorithms can be used to make computers analyze medical data accurately and generate multidimensional results. The pixel-wise classification of an image into logical related areas or volumes is called semantic segmentation. Different neural networks have their abilities in manipulating inputs and producing excellent results. Likewise, the deep learning technique U-Net outperforms the other network architectures for biomedical image segmentation. It is accurate and performs end-to-end semantic segmentation using an encoder and a decoder. U-Net is the popular approach for semantic medical image segmentation [7]. The first version of U-Net helped to crop and copy the feature map from the encoding unit to the decoding unit. It has significant advantages for segmentation tasks: first, the model allows the application of global location and context. Second, it gives good performance for the segmentation tasks with fewer training samples. It is using convolutional blocks and max-pooling in the encoder. The number of convolutional filters doubles at each level of the U-Net. The decoder uses de-convolution for up-sampling. While increasing the depth of a deep neural network, its accuracy may get saturated and degrades, and this degradation does not result from overfitting [23]. The residual learning makes the architecture less computationally complex, having the shortcut connection allowing the propagation of information without any degradation.

The deep Residual U-Net convolutional neural network uses a residual unit to extract discriminative features and overcomes the performance degradation by introducing a shortcut connection, which is an easier technique. Studies state that residual learning based neural network performs better than the sequential neural network [16]. This work applies data augmentation to generate synthetic data to enhance invariance property, which is shaped and illuminated. It applies online data augmentation that makes the number of augmented data equal to the number of total training data. The data augmentation includes flipping, shifting, rotation, and zooming. The analysis results state that shifting gives better improvement compared to rotation and flipping. This work applies post-processing to remove small areas of false-positives using connected components and applying thresholding. The publication has used a data dimension of $128 \times 128$ and achieved a DSC of 99.62 for 2D lung segmentation.

Multi-Scale Prediction Network gives predictions on multiple scales using single U-Net architecture. Using residual convolution blocks in a deep neural network solves gradient exploding and vanishing problems, providing the shortcut connection between input and output. The authors selected the lung CT scans from LUNA16 and NLST (National Lung Screening Trial) Dataset [18], having the criterion of selecting those CT scans that have interstitial lung disease and lung nodules attached on the lung wall [17].

The extension of the U-Net architecture using Recurrent Residual Convolutional Neural Networks called "R2U-Net" was evaluated in different fields of medical imaging [5]. The experimental results demonstrated better performance in 2D medical image segmentation. The residual units have an important role while training the deep architecture. The recurrent residual convolutional layers provide better feature representation for the segmentation tasks.

The recent 3D lung segmentation method – Extension of V-Net [6] is based on a modified version of the original V-Net [9]. It is using the max-pooling layer, in the beginning, to reduce the dimensions of the input scan from 256×512×512 to 64×128×128. This method randomly divide the available data of LUNA16 [3] into 700 scans and 188 scans for the validation of the neural network for 3D lung segmentation. The data augmentation techniques include spatial shifting and zooming along the depth axis.

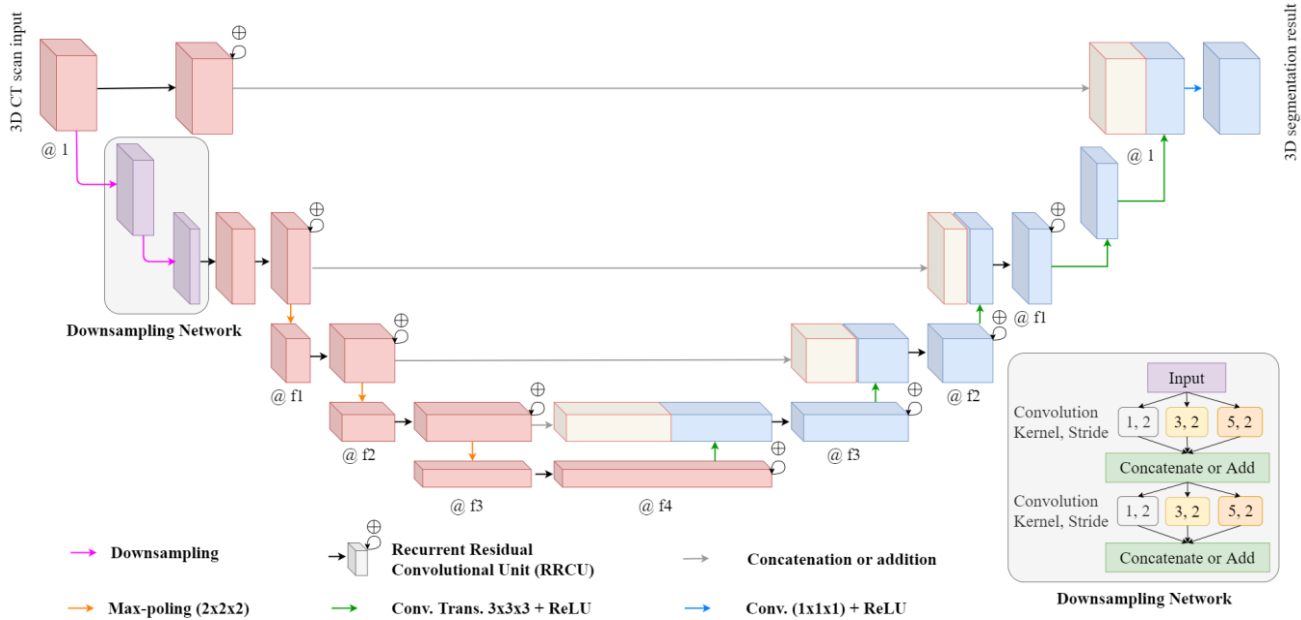

**FIGURE 1.** The overview of the proposed U-Net based R²U3D architecture for lung segmentation.

## III. PROPOSED FRAMEWORK

### A. R²U3D: Recurrent Residual 3D U-Net

Inspired by the concept of volumetric image segmentation, we have developed Recurrent Residual 3D U-Net (R²U3D) so that the proposed architecture can efficiently process volumetric data. The convolutional neural network learns by convolving over the multi-dimensional data. The convolution layer represents the spatial features, and the higher the dimension, the better the spatial features will be. Hence, a 3D convolutional neural network extracts the features according to 3D local and, ultimately, over the entire 3D volume.

The Recurrent Neural Network learns the spatial dependencies over multiple steps, and Residual Neural Network increases the propagation of 3D features. Considering these advantages, we have considered Recurrent and Residual Neural Networks based R²U3D as a base architecture and improved it by applying different neural network module – Squeeze-and-Excitation Residual module, loss functions – Soft-DSC and Exponential Logarithmic Loss, optimizers – Adam, proper learning strategies, and appropriate hyper-parameters. The analysis of deep neural networks becomes crucial while having hi-dimensional data. The proposed network architecture is illustrated in Fig. 1, consisting of a contracting path (left side) and an expansive path (right side). The left part is known as an encoder, and the right part is known as a decoder. The encoder consists of the down-sampling module, $3\times3\times3$ convolutions, Recurrent Residual Convolutional Unit (RRCU) (shown in Fig. 2), and the max-pooling layer followed by $1\times1\times1$ convolution. All of the convolutional units are followed by a Rectified Linear Unit (ReLU). Note that we down-sample the data in the beginning to avoid the hardware limitation. Instead of using the max-pooling layer, we are using Inception-like architecture for the down-sampling purpose. In particular, it has three convolutional layers with one filter with different kernel sizes. We either concatenate or add the output of each of them. This stretches the values (histogram) of the data and enhances the contrast. The decoder consists of $2\times2\times2$ up-convolution, RRCU, and the concatenation of the feature map from the encoder followed by $1\times1\times1$ convolution. The structure of **R²U3D** in terms of number of filters is $1 \rightarrow 1 \rightarrow 1 \rightarrow f1 \rightarrow f2 \rightarrow f3 \rightarrow f4 \rightarrow f3 \rightarrow f2 \rightarrow f1 \rightarrow 1 \rightarrow 1 \rightarrow 1$. We have applied the dilation in the encoder and the recurrent convolution unit. The sigmoid activation function follows the final layer.

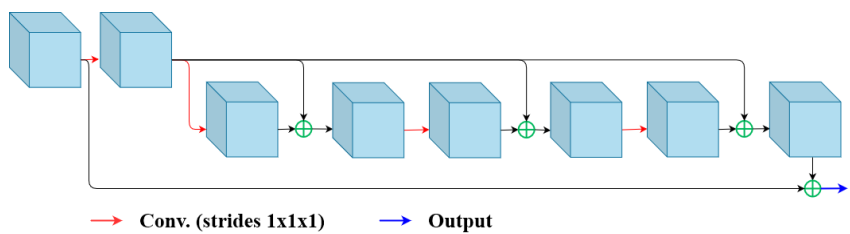

**FIGURE 2.** Recurrent Residual Convolutional Unit (RRCU) used in R²U3D. RRCU with a depth of 3 units.

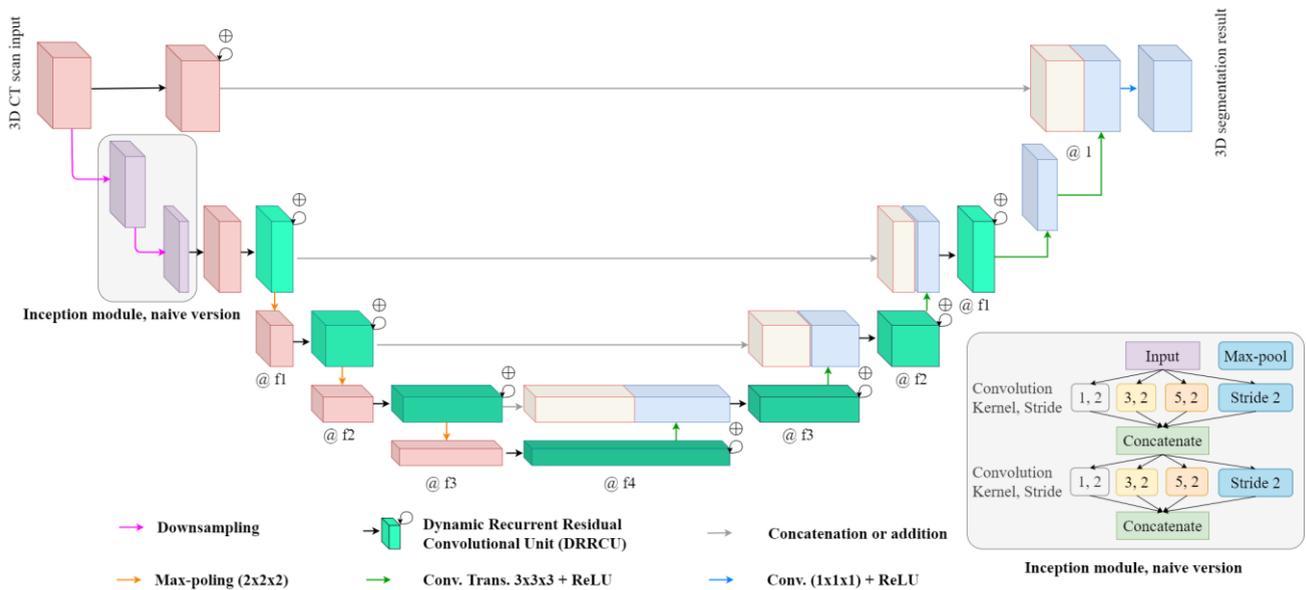

**FIGURE 3.** R²U3D architecture with Dynamic-Recurrent Residual Convolutional Unit (DRRCU).

The Recurrent Residual Convolutional Unit (RRCU) is an important representative module of our proposed architecture. The Recurrent convolutional unit accumulates the features for different depths and gives better feature representation. It ensures low-level feature accumulation on over the same levels of U-Net architecture.

### B. R²U3D Variants
#### 1) R²U3D WITH DEFAULT PARAMETERS
Fig. 1 shows a typical R²U3D architecture with the filters (f1, f2, f3, f4) = (40, 80, 160, 320). It has 20,306,691 parameters. The down-sampling network adds the outputs of convolutional layers. It uses Adam Optimizer with the learning rate of 0.001 and the loss function based on Dice Loss.

#### 2) R²U3D WITH DYNAMIC RECURRENT UNIT
The previous architecture has a less number of filters in each layer. It is necessary to increase the filters to make the deep neural network learn faster. To overcome this problem, we have introduced a Dynamic-Recurrent Residual Convolutional Unit (DRRCU). It primarily has a recurrent unit of different depth, with the Squeeze-and-Excitation Residual module in between. The purpose of applying different depth to the recurrent unit is to utilize the machine resources by eventually increasing the depth with approaching to the bottom layer of the architecture. That is less depth for the layers having a higher spatial dimension. This architecture has filters and depth of the recurrent unit {(f1, d1), (f2, d2), (f3, d3), (f4, d4))} = {(20, 1), (60, 2), (120, 3), (240, 4)}. It has 12,953,330 parameters. It uses Adam Optimizer with a learning rate of 0.001, and the loss function is the same as the previous architecture. We are using the Inception module, naïve version [12], with strides $2 \times 2 \times 2$ as a down-sampling network. It has three convolutional layers with one filter with different kernel sizes, along with a max-pooling layer. This architecture learns much faster than previous architectures. Fig. 3 shows the structure of R²U3D with the dynamic recurrent unit.

Inspired by Squeeze-and-Excitation Networks [10], DRRCU is having the Recurrent Neural Network followed by the Squeeze-and-Excitation Residual module. Fig. 4 represents the DRRCU at different depths. This unit helps to accumulate the low-level features for higher depth and utilizing the available machine resources. As shown in Fig. 5, the Squeeze-and-Excitation Residual module proposes the channel interdependencies and nonlinear interactions among the channels. It uses the global information of each of the channels and feeds it to two Fully Connected Networks (FCN). The first layer has the activation function ReLU, and the second has Sigmoid to normalize the output values from zero to one. The output of FCN is then multiplied with the input and generates the scaled input. The input is then added to the scaled input, followed by ReLU activation.

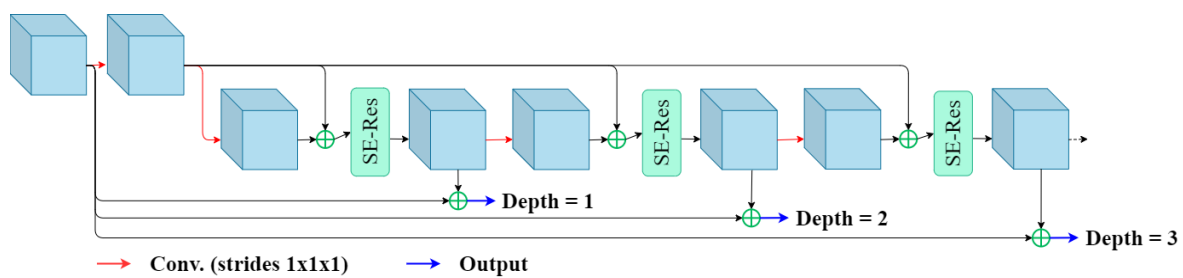

**FIGURE 4.** Dynamic-Recurrent Residual Convolutional Unit (DRRCU) with Squeeze-and-Excitation Residual module.

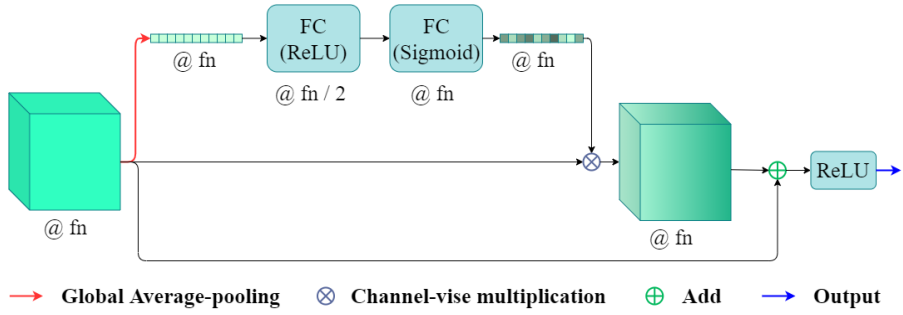

**FIGURE 5.** Squeeze-and-Excitation Residual module.

## IV. EXPERIMENTAL SETUP

We have implemented $R^2U3D$ using Keras deep learning library, with TensorFlow [35] as backend and Nvidia GeForce RTX 2080 Ti having 12 GB graphics memory.

### A. Dataset Details

We used publicly available datasets – LUng Nodule Analysis 2016 (LUNA16) and VESsel SEgmentation in the Lung 2012 (VESSEL12) [3][4]. LUNA16 consists of 888, and VESSEL12 consists of 20 three-dimensional lung CT scans, along with the segmented ground truth. We have considered 876 CT scans in our analysis, out of which 700 CT scans are for training, and 176 CT scans are for testing. Some of the ground-truth of LUNA16 scans are having holes inside lung areas, and most of them represent nodules. We have used CT scans from LUNA16 for both training and testing, and CT scans from VESSEL12 for testing.

### B. Data Preparation and Training Settings

The spatial resolution of data is 256×512×512. The training and testing 3D CT scans vary in number of slices and hence, to down-sample and up-sample 3D CT scans into 256×512×512 dimension. We are repeating the slices if actual available slices are less than 256 and selecting equally over the available slices if they are more than 256. This is performed with the proper number of steps with over the z-axis so that, the sampled data preserve the actual shape. We are normalizing each CT scan in the range from 0 to 1. We are not applying any data augmentation technique. Our training strategy is based on the random selection of training data from a certain part of the dataset. We are selecting five scans randomly from the set of first 100 scans, train them for five epochs, and repeat the procedure for 500 iterations. This strategy helps the model to overcome the problem of overfitting, particularly over the local set of training data. We used the learning rate of 0.001 and 0.0001 for 400 and 100 iterations, respectively. We kept batch-size one according to the available computational resources.

## V. EVALUATION

### A. Performance Metrics and Loss Functions

For the evaluation, we adopt Dice Similarity Coefficient (*DSC*) as below.

$$DSC = \frac{2\sum_i^N p_i g_i}{\sum_i^N p_i + \sum_i^N g_i} \quad (1)$$

We consider *Soft-DSC* for the evaluation:

$$Soft\ DSC = \frac{2\sum_i^N p_i g_i}{\sum_i^N p_i^2 + \sum_i^N g_i^2} \quad (2)$$

where *N* is the number of voxels in each image, $p_i \in P$ is a voxel of predicted segmentation P, and $g_i \in G$ is a voxel of binary ground-truth G.

The Exponential Logarithmic Loss [11] is computed as below:

$$Loss_{ELL} = w_{DSC} Loss_{DSC} + w_{Cross} Loss_{Cross} \quad (3)$$

$$Loss_{DSC} = (-\ln(DSC))^{\gamma_{DSC}} \quad (4)$$

$$Loss_{Cross} = WCEL^{\gamma_{WCEL}} \quad (5)$$

The loss is calculated with the addition of *DSC* and Weighted Cross Entropy with Logits (*WCEL*) [36], with a ratio. *WCEL* is first applying the Logit function, which is the inverse of Sigmoid function, to the prediction, and then calculated the Weighted Cross Entropy. Logit is used to calculate the values before the Sigmoid Activation. Here, $w_{DSC} = 0.8$, $w_{WCEL} = 0.2$ and $\gamma_{DSC} = \gamma_{WCEL} = 0.3$.

### B. Results and Discussions

The deep neural network that is the extension of V-Net (Extended V-Net) [6] is trained with 700 CT scans of LUNA16. Whereas, we have considered the first 100 CT scans from LUNA16 for the training set, tested our architecture with VESSEL12, and compared the results with Extended V-Net and V-Net as shown in Table 1. The **R²U3D** (Default) and **R²U3D** (Dynamic) provides the results of Soft-DSC as 0.9881 and 0.9920, respectively. Since the training data are enough, the testing data should be more than 20 number of 3D CT scans. Testing on less data does not guarantee the generalization and may overfit the model even if we observe a good testing accuracy. Therefore, we have tested our architecture with the last 176 CT scans of LUNA16. The **R²U3D** (Default) and **R²U3D** (Dynamic) give Soft-DSC of 0.9831 and 0.9859, respectively.

While testing all the remaining 776 scans of LUNA16, **R²U3D** (Dynamic) gives the Soft-DSC 0.9828. While training one variant of **R²U3D** (Default) with the rest of 700 LUNA16 CT scans in seven phases having 100 CT scans each and 700 CT scans in total, it shows an increment in the result for testing 176 CT scans of LUNA16. By training in batch of 100: 1 – 100, 100 – 200, 200 – 300, 300 – 400, 400 – 500, 500 – 600, by applying transfer learning, and testing on 700 – 876 (176 scans), the DSC is 0.9813, 0.9815, 0.982, 0.982, 0.9818, 0.9822. After that, training scans 1 – 700, further improves the DSC to 0.9827. Thus, the test results on 176 scans improve with more training data.

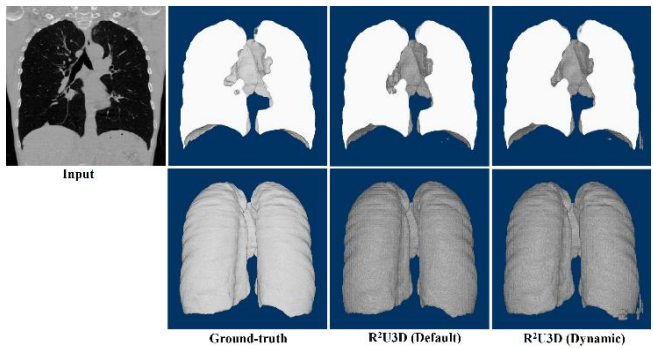

**FIGURE 6.** Visualization of 3D CT scan, ground-truth, and segmentation results using the proposed methods. The first row and second row show the dissected and entire lungs, respectively.

TABLE I
COMPARISON OF SOFT-DSC OF THE RESULTS FROM PROPOSED ARCHITECTURE WITH [6], FOR VESSEL 12 AND LUNA16 (LAST 176 SCANS) AS TESTING SETS. THE BEST RESULTS ARE MARKED AS BOLD-FACED.

|  | V-Net [10] | Extended V-Net [6] | R²U3D (Default) | R²U3D (Dynamic) |
|---|---|---|---|---|
| Training scans | 700 | 700 | 100 | 100 |
| Training iterations | 8400 | 8400 | 15650 | 12500 |
| Soft-DSC (VESSEL12) | 0.972 | 0.987 | 0.9881 | **0.9920** |
| Soft-DSC (LUNA16) | – | – | 0.9831 | 0.9859 |

### C. Applicability for COVID-19 Diagnosis
The deep learning model is trained on LUNA16. While applying it on 3D CT scans having COVID-19 infection, the segmentation predictions are imperfect and fail at infected areas. 3D CT scans in Fig. 7 are selected from COVID-19 CT Lung and Infection Segmentation Dataset [22]. Lung infectious diseases are having diverse patterns of infections and lesions. It is necessary to train the deep neural network on infected lungs based data to get better results on such testing dataset. This problem can be solved by training DNN on COVID-19 CT scans having lung masks, applying generative adversarial network (GAN), and generating synthetic data to help segment infected lungs.

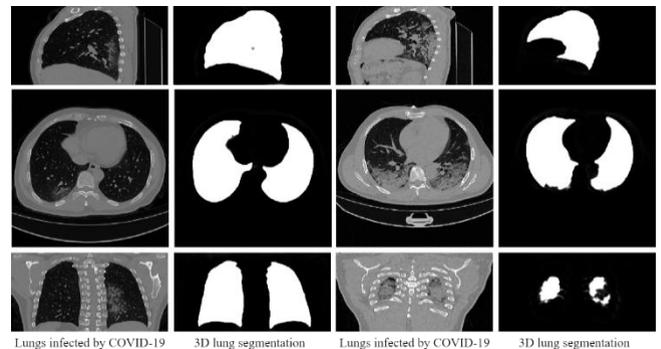

**FIGURE 7.** Visualization of lungs infected by COVID-19 the corresponding 3D lung segmentation.

## VI. CONCLUSIONS AND FUTURE WORK
In this paper, we have proposed the **R²U3D** network with its variants. We developed 3D deep neural network architecture of Dynamic-Recurrent Residual Convolutional Neural Network with a suitable down-sampling module and a Squeeze-and-Excitation Residual module, and trained with the Exponential Logarithmic Loss and Adam Optimizer. We trained our neural network on LUNA16 and tested it on both VESSEL12 and LUNA16 datasets. We have achieved better accuracies with less number of training data, and observed the improvement while training with the additional data.

Future work includes the application of medical imaging for diseases like lung cancer, chronic obstructive pulmonary disease (COPD), acute respiratory distress syndrome (ARDS), and pulmonary embolism (PE). These diseases damage the lungs and make imaging based tasks challenging. In addition, we aim to apply our methods for the segmentation of nodules from the segmented lungs, and classify them as malignant or benign. Furthermore, Coronavirus Disease 2019 (COVID-19) shows the regions having Ground-Glass Opacities (GGO) inside the lungs [13]. **R²U3D** can segment the lungs and be applied further to segment the GGO region from the segmented lungs using an appropriate dataset. The lungs infected by COVID-19 show diverse types of infection patterns other than GGO. It is essential to segment the infected lungs properly, and we plan to train the proposed deep neural network on COVID-19 based datasets. Thus we plan to employ data augmentation techniques that have shown promising results for this modality and design novel methods for robust lung segmentation.

## REFERENCES

1. American Cancer Society - Home, https://www.cancer.org/, last accessed 2020/11/04.
2. Peter Ellis and Rachel Vandermeer. "Delays in the diagnosis of lung cancer." Journal of thoracic disease 3, no. 3 (2011): 183.
3. LUNA16 - Home, https://luna16.grand-challenge.org/, last accessed 2020/11/04.
4. VESSEL12 - Home, https://vessel12.grand-challenge.org/, last accessed 2020/11/04.
5. Md Zahangir Alom, Mahmudul Hasan, Chris Yakopcic, Tarek M. Taha, and Vijayan K. Asari. " Recurrent Residual Convolutional Neural Network based on U-Net (R2U-Net) for Medical Image Segmentation." arXiv preprint arXiv:1802.06955 (2018).
6. Patrick Sousa, Adrian Galdran, Pedro Costa, and Aurélio Campilho. "Learning to Segment the Lung Volume from CT Scans Based on Semi-Automatic Ground-Truth." In 2019 IEEE 16th International Symposium on Biomedical Imaging (ISBI 2019), pp. 1202-1206. IEEE, 2019.
7. Olaf Ronneberger, Philipp Fischer, and Thomas Brox. "U-net: Convolutional networks for biomedical image segmentation." In International Conference on Medical image computing and computer-assisted intervention, pp. 234-241. Springer, Cham, 2015.
8. Md Zahangir Alom, Tarek M. Taha, Christopher Yakopcic, Stefan Westberg, Paheding Sidike, Mst Shamima Nasrin, Brian C. Van Esesn, Abdul A. S. Awwal, and Vijayan K. Asari. "The history began from alexnet: A comprehensive survey on deep learning approaches." arXiv preprint arXiv:1803.01164 (2018).
9. Fausto Milletari, Nassir Navab, and Seyed-Ahmad Ahmadi. "V-net: Fully convolutional neural networks for volumetric medical image segmentation." 2016 Fourth International Conference on 3D Vision (3DV). IEEE, 2016.
10. Jie Hu, Li Shen, and Gang Sun. "Squeeze-and-excitation networks." In Proceedings of the IEEE conference on computer vision and pattern recognition, pp. 7132-7141. 2018.
11. Ken CL Wong, Mehdi Moradi, Hui Tang, and Tanveer Syeda-Mahmood. "3D segmentation with exponential logarithmic loss for highly unbalanced object sizes." In International Conference on Medical Image Computing and Computer-Assisted Intervention, pp. 612-619. Springer, Cham, 2018.
12. Christian Szegedy, Wei Liu, Yangqing Jia, Pierre Sermanet, Scott Reed, Dragomir Anguelov, Dumitru Erhan, Vincent Vanhoucke, and Andrew Rabinovich. "Going deeper with convolutions." In Proceedings of the IEEE conference on computer vision and pattern recognition, pp. 1-9. 2015.
13. Shuchang Zhou, Yujin Wang, Tingting Zhu, and Liming Xia. "CT Features of Coronavirus Disease 2019 (COVID-19) Pneumonia in 62 Patients in Wuhan, China." American Journal of Roentgenology (2020): 1-8.
14. Ziabari, Amirkoushyar, Abbas Shirinifard, Matthew R. Eicholtz, David J. Solecki, and Derek C. Rose. "A Two-Tier Convolutional Neural Network for Combined Detection and Segmentation in Biological Imagery." In GlobalSIP, pp. 1-5. 2019.
15. De Nunzio, Giorgio, Eleonora Tommasi, Antonella Agrusti, Rosella Cataldo, Ivan De Mitri, Marco Favetta, Silvio Maglio et al. "Automatic lung segmentation in CT images with accurate handling of the hilar region." Journal of digital imaging 24, no. 1 (2011): 11-27.
16. Khanna, Anita, Narendra D. Londhe, S. Gupta, and Ashish Semwal. "A deep Residual U-Net convolutional neural network for automated lung segmentation in computed tomography images." Biocybernetics and Biomedical Engineering (2020).
17. Gu, Yuchong, Yaoming Lai, Peiliang Xie, Jun Wei, and Yao Lu. "Multi-Scale Prediction Network for Lung Segmentation." In 2019 IEEE 16th International Symposium on Biomedical Imaging (ISBI 2019), pp. 438-442. IEEE, 2019.
18. NLST - The Cancer Data Access System, https://cdas.cancer.gov/nlst/, last accessed 2020/11/04.
19. Xin, Yi, Gang Song, Maurizio Cereda, Stephen Kadlecek, Hooman Hamedani, Yunqing Jiang, Jennia Rajaei et al. "Semiautomatic segmentation of longitudinal computed tomography images in a rat model of lung injury by surfactant depletion." Journal of Applied Physiology 118, no. 3 (2015): 377-385.
20. Do, Synho, Kristen Salvaggio, Supriya Gupta, Mannudeep Kalra, Nabeel U. Ali, and Homer Pien. "Automated quantification of pneumothorax in CT." Computational and Mathematical methods in Medicine 2012 (2012).
21. Sahiner, Berkman, Aria Pezeshk, Lubomir M. Hadjiiski, Xiaosong Wang, Karen Drukker, Kenny H. Cha, Ronald M. Summers, and Maryellen L. Giger. "Deep learning in medical imaging and radiation therapy." Medical physics 46, no. 1 (2019): e1-e36.
22. Ma Jun, Ge Cheng, Wang Yixin, An Xingle, Gao Jiantao, Yu Ziqi, … He Jian. (2020). COVID-19 CT Lung and Infection Segmentation Dataset (Version Verson 1.0) [Data set]. Zenodo. http://doi.org/10.5281/zenodo.3757476
23. He, Kaiming, Xiangyu Zhang, Shaoqing Ren, and Jian Sun. "Deep residual learning for image recognition." In Proceedings of the IEEE conference on computer vision and pattern recognition, pp. 770-778. 2016.
24. Guan, Steven, Amir A. Khan, Siddhartha Sikdar, and Parag V. Chitnis. "Fully Dense UNet for 2-D Sparse Photoacoustic Tomography Artifact Removal." IEEE journal of biomedical and health informatics 24, no. 2 (2019): 568-576.
25. Park, S., L. C. Chu, E. K. Fishman, A. L. Yuille, B. Vogelstein, K. W. Kinzler, K. M. Horton et al. "Annotated normal CT data of the abdomen for deep learning: Challenges and strategies for implementation." Diagnostic and interventional imaging 101, no. 1 (2020): 35-44.
26. Tutt, Andrew. "An FDA for algorithms." Admin. L. Rev. 69 (2017): 83.
27. Singh, Amitojdeep, Sourya Sengupta, and Vasudevan Lakshminarayanan. "Explainable deep learning models in medical image analysis." arXiv preprint arXiv:2005.13799 (2020).
28. Meyes, Richard, Constantin Waubert de Puiseau, Andres Posada-Moreno, and Tobias Meisen. "Under the Hood of Neural Networks: Characterizing Learned Representations by Functional Neuron Populations and Network Ablations." arXiv preprint arXiv:2004.01254 (2020).
29. Kim, Mingyu, Jihye Yun, Yongwon Cho, Keewon Shin, Ryoungwoo Jang, Hyun-jin Bae, and Namkug Kim. "Deep learning in medical imaging." Neurospine 16, no. 4 (2019): 657.
30. Awad, Mariette, and Rahul Khanna. "Machine learning in action: examples." In Efficient Learning Machines, pp. 209-240. Apress, Berkeley, CA, 2015.
31. Zhou, Yongjin, Weijian Huang, Pei Dong, Yong Xia, and Shanshan Wang. "D-UNet: a dimension-fusion U shape network for chronic stroke lesion segmentation." IEEE/ACM transactions on computational biology and bioinformatics (2019).
32. Sun, Hui, Cheng Li, Boqiang Liu, Zaiyi Liu, Meiyun Wang, Hairong Zheng, David Dagan Feng, and Shanshan Wang. "AUNet: Attention-guided dense-upsampling networks for breast mass segmentation in whole mammograms." Physics in Medicine & Biology 65, no. 5 (2020): 055005.
33. Masood, Anum, Bin Sheng, Po Yang, Ping Li, Huating Li, Jinman Kim, and David Dagan Feng. "Automated decision support system for lung cancer detection and classification via enhanced RFCN with multilayer fusion RPN." IEEE Transactions on Industrial Informatics 16, no. 12 (2020): 7791-7801.
34. Qi, Kehan, Hao Yang, Cheng Li, Zaiyi Liu, Meiyun Wang, Qiegen Liu, and Shanshan Wang. "X-net: Brain stroke lesion segmentation based on depthwise separable convolution and long-range dependencies." In International Conference on Medical Image Computing and Computer-Assisted Intervention, pp. 247-255. Springer, Cham, 2019.
35. TensorFlow, https://www.tensorflow.org/, last accessed 2020/11/04.
36. https://www.tensorflow.org/api_docs/python/tf/nn/weighted_cross_entropy_with_logits, last accessed 2020/11/04.

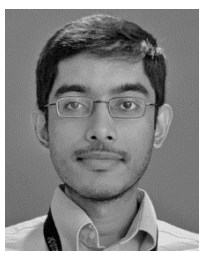

**DHAVAL D. KADIA** (Member, IEEE) received the B.E. degree in computer science from The Maharaja Sayajirao University of Baroda, India, in 2017, the M.S. degree in computer science from the University of Dayton, USA, in 2021.
He was a Research Intern at Defence Research and Development Organisation, India. He was a Teaching Assistant and Summer Fellow at the University of Dayton, USA. He was a member of Vision and Mixed Reality Laboratory, and Center of Excellence for Computational Intelligence and Machine Vision (Vision Lab) at the University of Dayton, USA. He is working as a Staff Research Associate at NCIRE - The Veterans Health Research Institute. His research interests include design and analysis of algorithms, image processing, computer vision, machine learning, deep learning, imaging, medical imaging, computer graphics, and mixed reality.
Mr. Kadia is a member of the Association for Computing Machinery (ACM). He received best paper award at International Conference on Computing, Analytics and Networks in 2017 and Graduate Student Summer Fellowship Award at the University of Dayton in 2019.

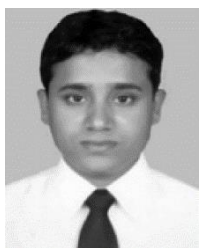

**MD ZAHANGIR ALOM** (Member, IEEE) received the B.Sc. degree (Hons.) in computer science and engineering from the University of Rajshahi, Bangladesh, in 2008, the M.Eng. degree in computer engineering from Chonbuk National University, South Korea, in 2012, and the Ph.D. degree in electrical and computer engineering from the University of Dayton, OH, USA, in 2018. He is working as a Bioinformatics Research Scientist at St. Jude Children's Research Hospital, USA. His research of interests include machine learning, deep learning, biomedical imaging, medical informatics, computational pathology, computer vision, and big data analytics.

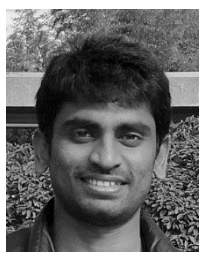

**RANGA BURADA** received his Bachelor's degree in electrical and communications engineering from JNTU-HYD, India, in 2011. He received his Master's degree in electrical engineering from the University of Dayton, USA, in 2015.
He is working as an Image Quality Engineer at Microsoft and pursuing a part-time Ph.D. His research interests are in developing Camera image quality metrics and image processing algorithms.

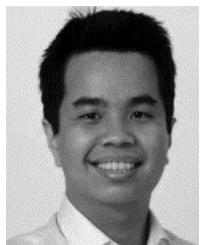

**TAM V. NGUYEN** (Senior Member, IEEE) received the B.S. degree in computer science from the University of Science, Vietnam, in 2005, the M.Eng. degree from Chonnam National University, South Korea, in 2009, and the Ph.D. degree from the National University of Singapore (NUS), in 2013. He is currently an Assistant Professor and the Director of Vision and Mixed Reality Laboratory, Department of Computer Science, University of Dayton. His research interests include computer vision, machine learning, mixed reality, and multimedia analysis.

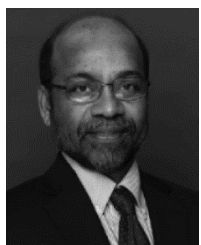

**VIJAYAN K. ASARI** (Senior Member, IEEE) is the University of Dayton Ohio Research Scholars Endowed Chair in Wide Area Surveillance and a Professor with the Department of Electrical and Computer Engineering. He is also the Director of the Center of Excellence for Computational Intelligence and Machine Vision (Vision Lab). He has received many awards for his teaching, research, and technical leadership, including the Vision Award for Excellence, in August 2017, the Sigma Xi George B. Noland Award, in April 2016, and the Outstanding Engineers and Scientists Award for Technical Leadership, in April 2015. He has published, and coauthored with his graduate students and colleagues, more than 700 research articles, including more than 110 peer-reviewed journal articles, in the areas of image processing, computer vision, pattern recognition, machine learning, deep learning and high performance digital system architecture design. He is a Fellow of the SPIE. He has co-organized several IEEE and SPIE conferences and workshops.